\documentclass[floatfix, twocolumn, prb]{revtex4}

\pdfoutput=1
\usepackage{color}
\usepackage{multirow}
\usepackage{epsfig}
\usepackage{graphicx}
\usepackage{amsmath}
\usepackage{amssymb}
\usepackage{bm}
\usepackage{dcolumn}
\usepackage{braket}
\usepackage{bbold}
\usepackage{ulem}
\usepackage{soul}
\usepackage{rotating}
\usepackage{tikz,pgfplots,gensymb}
\usepackage[hidelinks]{hyperref}
\usepackage{xcolor}
\usepackage{pifont}
\usepackage{array}
\usepackage{caption}
\usepackage{subcaption}
\pgfplotsset{compat=1.8}
\usetikzlibrary{pgfplots.groupplots}

\hypersetup{
    colorlinks,
    linkcolor={red!50!black},
    citecolor={blue!50!black},
    urlcolor={blue!80!black}
}

\begin{document}

\title{Piezospintronic effect in antiperovskite Mn$_3$GaN}

\author{Ilias Samathrakis}
\author{Hongbin Zhang}
\email[corresp.\ author: ]{hzhang@tmm.tu-darmstadt.de}

\affiliation{Institute of Materials Science, TU Darmstadt, 64287 Darmstadt, Germany}

\begin{abstract}
Based on first-principles calculations, we investigated the topological transport properties of Mn$_3$GaN with 
coplanar noncollinear magnetic structures. The intrinsic anomalous Hall conductivity (IAHC) displays a significant dependence with respect to the in-plane magnetization direction between the $\Gamma_{5g}$ and $\Gamma_{4g}$ magnetic configurations, 
where large anomalous Nernst effect (ANE) can be induced by tailoring the magnetization direction. Moreover, we observed strong piezospintronic effect in Mn$_3$GaN, where large IAHC can be induced by moderate epitaxial strain. Symmetry analysis reveals that for both cases, the nonzero IAHC is originated from the spin-orbit coupling instead of the noncollinear magnetic configurations.
\end{abstract}

\maketitle

\section{\label{INTRO}Introduction}
\par Antiferromagnetic (AFM) spintronics has recently drawn intensive attention due to its advantages over the ferromagnetic (FM) counterpart.~\cite{Jungwirth:2016} For instance, due to the vanishing net magnetization, the AFM magnets have negligible stray fields on the neighbouring structures. Moreover, the AFM devices can operate in the THz range,~\cite{Jungwirth:2016, Gomonay:2014, VZelezny:2017, Smejkal:2017} which is much faster than the ferromagnetic cases. It is therefore desirable to use AFM magnets, which are more abundant than the ferromagnetic materials, in spintronics applications. In general, AFM spintronics targets at either generating spin current or controlling the magnetization by means of electric voltage or current.~\cite{Kimura:2003} For instance, spin-orbit torque, which was firstly detected by Chernyshov~\cite{Chernyshov:2009} originated from the Edelstein effect~\cite{Edelstein:1990} due to spin-orbit coupling (SOC), has been demonstrated to be able to manipulate the AFM magnetization direction by electric currents in CuMnAs.~\cite{Wadley:2013} This enables the possibility to engineer magnetic memory or logic units using AFM magnets in future spintronic applications.~\cite{Zelezny:2014}

\par Regarding the generation of spin currents, magnetic materials with noncollinear magnetic structure is of particular interest. 
The Mn$_3$X systems with X = Sn, Ge, Ir, Ga, Rh, and Pt  have been proposed to host large intrinsic anomalous Hall conductivity (IAHC) and spin Hall conductivity (SHC),~\cite{Chen:2014, Kubler:2014,Nayak:2016, Guo:2017, YZhang:2017} which has been confirmed by recent experiments on Mn$_3$Sn and Mn$_3$Ge.~\cite{Nakatsuji:2015, Kiyohara:2016} Such nontrivial topological transport properties have been observed in many other systems such as Mn$_5$Si$_3$,~\cite{Surgers:2017, Surgers:2018} Pr$_2$Ir$_2$O$_7$,~\cite{Machida:2007,Udagawa:2013} and Nd$_2$Mo$_2$O$_7$.~\cite{Taguchi:2001} One interesting question is whether the nonzero IAHC or SHC is caused by the noncollinear magnetic structure, as argued in Ref.~\onlinecite{Nayak:2016} or the SOC. The former mechanism leads to the so-called topological Hall effect (THE), which can get quantized as proposed in K$_{0.5}$RhO$_2$.~\cite{Zhou:2016} 

\par In this work, we demonstrated that magnetic antiperovskite materials with noncollinear magnetic structures are interesting for AFM spintronic applications. Taking Mn$_3$GaN as an example, our density function theory (DFT) calculations revealed two ways of inducing nonzero IAHC. Firstly, we observed that IAHC has a strong dependence with respect to the rotation of the magnetization direction, leading to finite IAHC and large anomalous Nernst effect. Secondly, nonzero IAHC can also be induced utilizing the piezomagnetic effect,~\cite{Lukashev:2008} hereafter dubbed as piezospintronic effect. It is demonstrated explicitly that for both cases, the IAHC is not caused by the noncollinear magnetic structure but by the atomic SOC, due to the coplanar nature of the magnetic configurations. 

\section{Numerical details}

\par Our calculations are performed using the projector augmented wave (PAW) method as implemented in the VASP code.~\cite{Kresse:1993} The exchange-correlation functional is approximated in the generalized gradient approximation (GGA) as parameterized by Perdew-Burke-Ernzerhof.~\cite{Perdew:1996} A $\Gamma$-centered $12 \times 12 \times 12$ k-mesh and a 500~eV energy cutoff are used in all our calculations to guarantee good convergence. The valence electron configurations of  $3d^{6}4s^{1}$, $3d^{10}4s^{2}4p^{1}$ and $2s^{2}2p^{3}$ are taken into account for Mn, Ga, and N atoms, respectively.

\par The equilibrium lattice parameters of the Mn$_3$GaN are obtained by performing relaxation to the system. Our results reveal a cubic ground state with $a=3.864$~\AA~and magnetic moment of 2.39~$\mu_B$. This is consistent with the experimentally reported lattice constant value of $3.898$\AA, found in Ref.~\onlinecite{Bertaut:1968} as well as the calculated magnetic moment of 2.40~$\mu_B$.~\cite{Lukashev:2008} The applied biaxial strain is performed by imposing values within $\pm4\%$ of the lattice parameter $c$ to the equilibrium lattice parameter $c$. Subsequently, several $a$ and $b$ are tested and the optimal ones are extracted by using a polynomial fitting to the energy curve obtained for each of the imposed $c$ values. 

\par To evaluate the IAHC, the DFT Bloch wave functions are projected onto the maximally localised Wannier functions (MLFW) following Ref.~\onlinecite{Mostofi:2008}. In our work, $s$, $p$ and $d$ orbitals for Mn and Ga atoms, and $s$ and $p$ for N atoms are projected, summing up to $80$ MLWFs in total. In this paper, the IAHC was evaluated by integrating the Berry curvature according to the formula:~\cite{Xiao:2010}

\begin{align}
& \sigma_{\alpha \beta} = - \frac{e^2}{\hbar} \int \frac{d\mathbf{k}}{\left( 2\pi \right)^3} \sum_{n} f \big[ \epsilon \left( \mathbf{k} \right) -\mu \big] \Omega_{n,\alpha \beta} \left( \mathbf{k} \right) \notag \\
& \Omega_{n,\alpha \beta}\left( \mathbf{k} \right) = -2Im \sum_{m \neq n} \frac{\braket{\mathbf{k}n|v_{\alpha}|\mathbf{k}m}\braket{\mathbf{k}m|v_{\beta}|\mathbf{k}n}}{\big[ \epsilon_m\left( \mathbf{k} \right) - \epsilon_n\left( \mathbf{k} \right) \big]^2} \notag,
\end{align}
\newline

\noindent with $f\big[ \epsilon \left(\mathbf{k}\right) - \mu \big]$ being the Fermi distribution function, $\mu$ the fermi energy, $n$ ($m$) the occupied (empty) Bloch band, $\epsilon_{n}\left(\mathbf{k}\right)$ ($\epsilon_{m}\left(\mathbf{k}\right)$) their energy eigenvalues and $v_{\alpha}$ ($v_{\beta}$) the velocity operator. The integration was performed on a $240 \times 240 \times 240$ kmesh using Wannier90.~\cite{Mostofi:2008}

\section{\label{RES}Results and Discussion}

\subsection{In-plane rotation}

\par
\begin{figure*}
        \centering
 \includegraphics[width=0.7\textwidth]{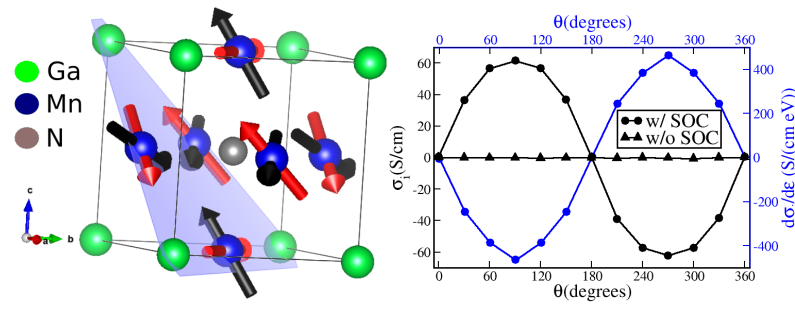}
\caption{\small (a): In-plane rotation of the magnetic moments of all 3 Mn atoms. The red arrows correspond to the starting angle of $\theta=0\degree$ ($\Gamma_{5g}$), whereas the black ones illustrate the $\theta=90\degree$ ($\Gamma_{4g}$). (b): IAHC (black) and ANC (blue) as a function of angle of in-plane rotation in the presence (circles) and absence (triangles) of SOC.}
\label{fig1}
\end{figure*}

 The magnetic ground state of Mn$_3$GaN is found to be of the $\Gamma_{5g}$-type (Fig.~\ref{fig1}a), {\it i.e.}, 
with the magnetic moments of Mn atoms lying in the (111)-plane and aligned along the face-diagonal directions, which is 0.11 eV lower in energy than the ferromagnetic configuration. Moreover, considering SOC, the $\Gamma_{5g}$ is about 0.5~meV per unit cell lower than the $\Gamma_{4g}$ configuration where the magnetic moments are rotated by 90$^\circ$ coherently in the (111)-plane (see Fig.~\ref{fig1}a). Actually, the energy versus the rotation angle of magnetic moments in the (111)-plane shows a typical sinusoidal behaviour. We note that the in-plane rotation of the Mn moments is energetically more favorable than the out-of-plane tilting as considered for Mn$_3$Ir.~\cite{Chen:2014} For instance, explicit calculations reveal that an out-of-plane canting of 2$^\circ$ degrees costs 1.1 meV for Mn$_3$GaN. On the other hand, the magnetocrystalline anisotropy within the (111)-plane for Mn$_3$GaN is enhanced compared to that of Mn$_3$NiN, which is only about 0.15 meV. This can be attributed to the significant atomic SOC strength of Ga atoms compared to Ni. Nevertheless, it is realized that the magnetocrystalline anisotropy for Mn$_3$Sn is only about 0.026 meV, though Sn is heavier, thus with larger SOC strength than Ga. We suspect that such a difference can be attributed to different crystalline structures, which requires further detailed investigation.

\par Interestingly, for the $\Gamma_{5g}$ and $\Gamma_{4g}$ magnetic configurations with competitive energies, the IAHC shows
quite distinct behavior, as shown in Fig.~\ref{fig1}b. Firstly, without SOC, the IAHC is zero, indicating that there exists no
topological Hall effect caused by the coplanar noncollinear magnetic structure. Secondly, the IAHC is antisymmetric with
respect to the magnetization direction, {\it i.e.}, $\sigma(\mathbf{M})=-\sigma(-\mathbf{M})$, though the total energies are
the same for two magnetic configurations where the magnetization direction is rotated by 180$^\circ$. For the $\Gamma_{4g}$ state, all three components of the IAHC are nonzero with the same magnitude after considering SOC. As a matter of fact, the finite IAHC can be induced by a finite rotation away from the $\Gamma_{5g}$ configuration, reaching its maximum about 62 S/cm (for each componet) in the $\Gamma_{4g}$ state (Fig.~\ref{fig1}b). In this regard, SOC plays an essential role in inducing finite IAHC in Mn$_3$GaN.
 
\par Symmetry plays an essential role to understand the IAHC of Mn$_3$GaN. In the absence of the SOC (and for coplanar magnetic configurations), the system is invariant under the $R_ST$ symmetry (effective $T$ symmetry), which is the combination of the time reversal symmetry ($T$) and spin the rotation ($R_S$). As a result, the Berry curvature, that transforms according to $\Omega^a \left(k_x,k_y,k_z\right)=-\Omega^a \left(k_x,k_y,k_z\right)$, vanishes after integrating over the whole Brillouin zone,~\cite{Suzuki:2017} leading to zero topological Hall effect (see the red curve of Fig.~\ref{fig1}b). In the presence of SOC though (and also for coplanar magnetic configurations), the effective $R_ST$ symmetry is broken and the presence of the IAHC depends on the symmetries each magnetic configuration possesses. 

For the $\Gamma_{5g}$ state with magnetic space group $R\bar{3}m$ (166.97), the $M_{1\bar{1}0}$ (and similar) mirror symmetries are preserved. Correspondingly, the Berry curvature transforms according to $\Omega_{xy}\left(k_x,k_y,k_z\right)=-\Omega_{xy}\left(-k_y,-k_x,-k_z\right)$, leading to vanishing IAHC after integrating over the whole Brillouin zone. For magnetic configurations deviated from $\Gamma_{5g}$ by a finite angle, the resulting magnetic space group is $R\bar{3}$ (148.17) and there is no symmetry operation which prohibits finite IAHC. Moreover, as there is still three fold rotational symmetry by the (111)-axis, three components of the IAHC tensor are of the same magnitude. The results based on the symmetry analysis are fully consistent with our numerical values of IAHC, which is further confirmed by explicitly constructed conductivity tensors following Ref.~\onlinecite{Jakub} for the magnetic space groups $R\bar{3}m$, $R\bar{3}m'$ and $R\bar{3}$ (Table~\ref{ahc-tensor}). Similar behavior is observed in Mn$_3$GaN by the others~\cite{Gurung:2019} and in Mn$_3$NiN.~\cite{Boldrin:2019}

\begin{table}
\centering
\caption{AHC tensor for different magnetic configurations. The symbols as well as the numbers of the magnetic space groups correspond to the Belov-Neronova-Smirnova (BNS) settings. The following notation of the IAHC components has been used: $\sigma_{yz}=\sigma_x$, $\sigma_{zx}=\sigma_y$ and $\sigma_{xy}=\sigma_z$.}
\begin{tabular}{|c|c|c|}
\hline
\bf{Magnetic} &\bf{Magnetic} & \bf{AHC} \\
\bf{configuration} & \bf{space group} & \bf{tensor} \\
\hline \hline
$\Gamma_{5g}$ & $R\bar{3}m$ (166.97) & $\left[ \begin{array}{ccc} 0 & 0 & 0 \\ 0 & 0 & 0 \\ 0 & 0 & 0 \end{array}\right]$ \\
\hline
$\Gamma_{4g}$ & $R\bar{3}m'$ (166.101) & $\left[ \begin{array}{ccc} 0 & \sigma_{xy} & -\sigma_{xy} \\ -\sigma_{xy} & 0 & \sigma_{xy} \\ \sigma_{xy} & -\sigma_{xy} & 0 \end{array}\right]$ \\
\hline
$\theta\neq0,90$ & $R\bar{3}$ (148.17) & $\left[ \begin{array}{ccc} 0 & \sigma_{xy} & -\sigma_{xy} \\ -\sigma_{xy} & 0 & \sigma_{xy} \\ \sigma_{xy} & -\sigma_{xy} & 0 \end{array}\right]$ \\
\hline
Distorted $\Gamma_{5g}$ & $C2/m$ (12.58) & $\left[ \begin{array}{ccc} 0 & 0 & -\sigma_{zx} \\ 0 & 0 & -\sigma_{zx} \\ \sigma_{zx} & \sigma_{zx} & 0 \end{array}\right]$ \\
\hline
Distorted $\Gamma_{4g}$ & $C2'/m'$ (12.62) & $\left[ \begin{array}{ccc} 0 & \sigma_{xy} & -\sigma_{zx} \\ -\sigma_{xy} & 0 & \sigma_{zx} \\ \sigma_{zx} & -\sigma_{zx} & 0 \end{array}\right]$ \\
\hline
\end{tabular}
\label{ahc-tensor}
\end{table}

\par The IAHC tensor depends on the geometry of the measurements. The symmetry imposed tensor shape of the IAHC displayed in Tab.~\ref{ahc-tensor} assumes the standard Cartesian coordinates $a||\big[100 \big]$, $b||\big[010 \big]$ and $c||\big[ 001 \big]$ (hereafter referred as the ``first basis"). However, different directions can be chosen in order to simplify the form of the IAHC tensor and make comparisons with the experiment more straightforward. As discussed above, the IAHC has three-fold rotational symmetry by the [111]-axis, leading to an IAHC tensor which is parallel to the [111]-axis. We calculated the IAHC tensor in the corresponding hexagonal geometry, {\it i.e.}, $a||\big[ 1\bar{1}0 \big]$, $b||\big[ 10\bar{1} \big]$ and $c||\big[ 111 \big]$ axes. It is observed that there is only one nonzero component for the IAHC when the magnetic configuration is deviated from $\Gamma_{5g}$. For instance, a value of IAHC as large as 102 S/cm is obtained for the $z||[111]$ component in the $\Gamma_{4g}$ arrangement, corresponding to $\sqrt{3}$ times the $\sigma_{x/y/z}$ in the first basis, being consistent with the value of 96.3 S/cm calculated in Ref.~\onlinecite{Huyen:2019}. 

\par The IAHC, tunable by changing the magnetization direction, leads to significant anomalous Nernst effect. Based on the Mott relation,\cite{Ashcroft:2010} each of the anomalous Nernst coefficient (ANC) components can be expressed as the energy derivative of the corresponding anomalous Hall conductivity component $\frac{d\sigma_i}{d\epsilon}$. It is observed that all ANC components vanish in the $\Gamma_{5g}$ ground state, as IAHC is zero, regulated by symmetry (see Fig.~\ref{fig1}b). While the angle of rotation increases, the absolute magnitude of each component increases as well, till it peaks at $\approx -500 \frac{S}{cm \cdot eV}$ for the $\Gamma_{4g} \left( \theta = 90 \right)$ state (see Fig.~\ref{fig1}b). The maximum value is comparable to the $-527 \frac{S}{cm \cdot eV}$ reported in Mn$_3$Sn by Ref.~\onlinecite{Guo:2017}, suggesting that significant ANC can be induced by tailoring the magnetization directions in Mn$_3$GaN. We note that the ANC can be further enhanced to $\approx -9000 \frac{S}{cm \cdot eV}$ (per component) for hole-doped cases with 0.3 holes per unit cell.

\subsection{Piezospintronic effect}

\begin{figure*}
 \includegraphics[width=0.7\textwidth]{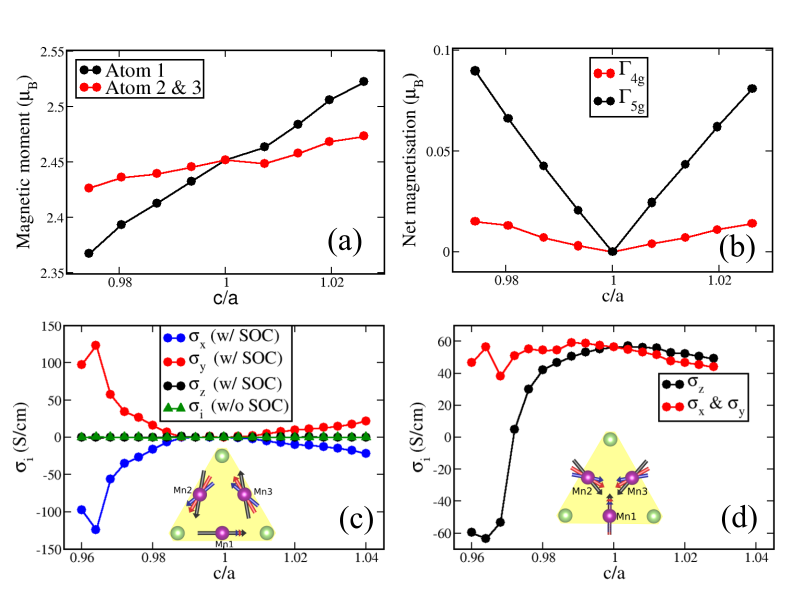}
\caption{\small (a): Magnetic moments of all Mn atoms as a function of strain in both $\Gamma_{5g}$ and $\Gamma_{4g}$. 
(b): Net magnetisation as a function of strain in $\Gamma_{5g}$ and $\Gamma_{4g}$ states. (c): IAHC (circles) and THE (triangles) as a function of strain in $\Gamma_{5g}$ state. Inset: Local magnetic moments for the unstrained $\Gamma_{5g}$ state (red) and strained (tensile with black and compressive with blue). (d): The components of IAHC as a function of strain in $\Gamma_{4g}$ state. Inset: Local magnetic moments for the unstrained $\Gamma_{4g}$ state (red) and strained (tensile with black and compressive with blue)}
\label{fig2}
\end{figure*}

 Detailed constrained calculations reveal the presence of piezomagnetic response in Mn$_3$GaN in both $\Gamma_{5g}$ and $\Gamma_{4g}$ states. The change in the magnitude of the local magnetic moments influences the induced net magnetisation in the $\Gamma_{5g}$ state. It is observed that, upon applied strain, the magnetic moments of all three Mn atoms stay within the (111) plane of the tetragonally distorted unit cell. However, the moments of the two Mn atoms on the side faces deviate from the $\Gamma_{5g}$ arrangement  by rotating by the $\big[111\big]$-axis, as sketched in the inset of Fig.~\ref{fig2}c. As a result, the net magnetisation takes values as large as -0.04 $\mu_B$ (0.03 $\mu_B$) for $c/a = 0.99$ ($c/a=1.01$) (see Fig.~\ref{fig2}b), being consistent with previous studies, the results of which ($\approx$ 0.04 $\mu_B$)~\cite{Lukashev:2008,Zemen:2015} match our outcome, suggesting that the magnetism has a very strong itinerant nature.

Regarding the $\Gamma_{4g}$ state, it is observed that, upon applied strain, the magnetic moments of all three Mn atoms stay within the (111)-plane of the tetragonally distorted unit cell, similarly to the $\Gamma_{5g}$ case. However, the moments of the two Mn atoms on the side faces deviate from the $\Gamma_{4g}$ arrangement, both canting towards (against) the Mn atom at the basal plane, when tensile (compressive) strain is applied, as illustrated in the inset of Fig.~\ref{fig2}d. A smaller tilt of -0.5$^\circ$ (0.5$^\circ$) as a response to -1\% (1\%) strain is observed, leading to a smaller induced net magnetisation of the order of $\mp$0.007 $\mu_B$ for $c/a = 0.99$ and $c/a=1.01$ respectively (see Fig.~\ref{fig2}b).

\par The reduction of the crystalline and magnetic symmetries leads to changes in the anomalous Hall conductivity. More specifically, a difference of around 5 S/cm in the absolute value of the $\sigma_{yz}=\sigma_{x}$ and the $\sigma_{zx}=\sigma_{y}$ components is observed as a response to $\pm1\%$ applied strain in the $\Gamma_{5g}$ state (see Fig.~\ref{fig2}c). This small deviation drastically alters when a larger strain is applied. In particular, an increase of 125 S/cm in the absolute value of same components is observed as a response to 3.5\% applied compressive strain. The latter values of these two components are comparable to the 133 S/cm, calculated for Mn$_3$GaN in Ref~\onlinecite{YZhang:2017}, demonstrating that large IAHC values can be induced by applying strain to a system with negligible values in its ground state, dubbed piezospintronic effect. A significantly smaller response to strain is observed in the $\Gamma_{4g}$ state, where a 3.5\% applied compressive strain  alters the value of $\sigma_{x}$ and $\sigma_{y}$ components by approximately 20 S/cm (from 40 S/cm to 60 S/cm) (see Fig.~\ref{fig2}d). The piezospintronic effect has been very recently observed in Mn$_3$NiN (another member of the antiperovskite family), where changes of approximately 100 S/cm (from 200 S/cm to 300 S/cm) have been calculated as a response to 1.5\% applied compressive strain.~\cite{Boldrin:2019}

\par The calculated IAHC is determined by the magnetic space group and the underlying possessed symmetries of the unit cell. The applied strain (either tensile or compressive) in the pure $\Gamma_{5g}$ state alters the magnetic space group from $R\bar{3}m$ (166.97) to $C2/m$ (12.58). Consequently, the symmetries possessed in the latter space group, force $\sigma_x$ and the $\sigma_y$ to have finite values, being the additive inverse of each other and leading to the symmetrical curve displayed in Fig.~\ref{fig2}c. On the other hand, the magnetic space group changes from $R\bar{3}m'$ (166.101) to $C2'/m'$ (12.62) under applying strain, in the $\Gamma_{4g}$ configuration, leading to identical $\sigma_{x}$ and $\sigma_{y}$ components (shown in Fig.~\ref{fig2}d). Our results agree with the symmetry imposed form of the AHC tensor for the magnetic space groups $C2/m$ and $C2'm'$, which were constructed using Ref~\onlinecite{Jakub} and they are shown in Tab.~\ref{ahc-tensor}. Since the AHC depends on the SOC, finite values are expected in different compounds of the antiperovskite family M$_3$AN. Recent realisations have demonstrated the presence of nonzero values in Mn$_3$NiN.~\cite{Boldrin:2019}

\section{\label{CONC}Conclusion}

\par Our first-principles calculations on the antiperovskite Mn$_3$GaN reveal that the IAHC is driven by SOC not by the coplanar noncollinear magnetic structure, and it is further demonstrated that the IAHC can be manipulated by in-plane rotation of the magnetic moments or by imposing biaxial strain. For the $\Gamma_{4g}$ configuration, the nonzero IAHC leads to significant anomalous Nernst effect. The (111)-heterostructures show only one nonzero IAHC component, consistent with detailed symmetry analysis, which can be tailored by in-plane magnetic fields. It is observed that compressive strain can induce more significant change of the IAHC, due to the larger deviation of the magnetic moments from the $\Gamma_{5g}$ state. In practice, we suspect that the IAHC can be tuned by applying strain imposed by BaTiO$_3$, as discussed in Ref~\onlinecite{Shao:2019}. Additionally, the spin hall effect and the magneto-optic Kerr effect (MOKE) consist two additional effects to be investigated in the magnetic antiperovskite family, leading to interesting spintronics applications. 

\begin{center}
\small \textbf{ACKNOWLEDGMENTS}
\end{center}

This work was financially supported by the Deutsche Forschungsgemeinschaft (DFG) via the priority programme SPP 1666 and the calculations were conducted on the Lichtenberg high performance computer of the TU Darmstadt. Additionally, we thank Harish Kumar Singh and Nuno Fortunato for valuable discussions.

%\newpage

%\bibliographystyle{ieeetr}
%merlin.mbs aipauth4-1.bst 2010-07-25 4.21a (PWD, AO, DPC) hacked
%Control: key (0)
%Control: author (9) reversed initials
%Control: editor formatted (0) differently from author
%Control: production of article title (-1) disabled
%Control: page (0) single
%Control: year (1) truncated
%Control: production of eprint (0) enabled
%

\end{document}